\documentclass[epj]{webofc}
\usepackage[varg]{txfonts}   
\woctitle{the Time Machine Factory 2012}
\begin{document}
\vspace{-4em}  
\title{Elementary cycles of time}
%
%
\author{Donatello Dolce \inst{1}\fnsep\thanks{\email{donatello.dolce@unicam.it}}
}

\institute{ CoEPP, The University of Melbourne, Australia \&
          Comerino University, Italy.
                    }
\vspace{-1em}  
\abstract{
Elementary particles, i.e. the basic constituents of nature, are characterized by quantum recurrences in time.  The flow of time of every physical system can be therefore decomposed in elementary cycles of time.  This allows us to enforce the local nature of relativistic time, yielding interesting unified descriptions of fundamental aspects of modern physics, as shown in recent publications and reviewed in \cite{Dolce:spatimecyles}. Every particle can be regarded as a reference clock with time resolution of the order of its Compton time, typically many orders of magnitude more accurate than the atomic clocks.  Here we summarize basic conceptual aspects of the resulting relational interpretation of the relativistic time flow.   
%
}
\maketitle
\vspace{-1.5em}  
\section{Introduction}
\label{intro}

In recent publications \cite{Dolce:spatimecyles,Dolce:2009ce,Dolce:tune,Dolce:AdSCFT} we have proposed a formalization of relativistic Quantum Mechanics (QM) in terms of relativistic Elementary Cycles (ECs). This description can be justified, for instance, by the wave-particle duality introduced by de Broglie in terms of a ``periodic phenomenon'' attributed to every elementary particle \cite{Broglie:1924}.  In a ``periodic phenomenon'' the space-time coordinates enter as  angular variables, e.g. as in waves and phasors. By assuming ECs in space-time, the recurrence of elementary particle can be implemented directly at the level of space-time geometrodynamics. Technically this is realized  in terms of compact space-time dimensions with periodic boundary conditions \cite{Dolce:2009ce,Dolce:tune,Dolce:AdSCFT}.  The resulting description provides elegant solutions of central problems of modern physics such as a unified semiclassical geometrodynamical description: of quantum and classical mechanics \cite{Dolce:2009ce}, of gauge and gravitational interaction \cite{Dolce:tune}, as well as an intuitive interpretation of Maldacena's conjecture \cite{Dolce:AdSCFT}. 

\vspace{-1em}  
\section{Elementary cycles}
\label{sec-1}

In physics every system can be described in terms of a set of elementary particles and their interaction schemes.  The Standard Model predicts with astonishing precision the physical properties of the experimentally observed matter  in terms of basic constituents: leptons (electron, muon, tau, and their neutrinos) and quarks (up, down, charm, strange, top and bottom), constituting proper matter particles; vector bosons (photon, $W^{\pm}$, $Z$, and gluons) which, together with gravity, describe the possible interactions of the matter particles; and recently the experimental evidences from LHC of the Higgs boson, generating the mass of the the elementary particles. 

These elementary particles are characterized by different properties such  as mass, charges, and spin, determining their relativistic evolutions in space-time. Interactions and the consequent dynamics of a single particle are denoted by corresponding relativistic variations of its energy $\bar E(\bar{\mathbf{p}})$ and momentum $\bar{\mathbf{p}}$. For a free (isolated) particle the four-momentum $\bar p_{\mu}=\{\bar E ,-\mathbf{\bar{p}}\}$ is constant; it has uniform motion  (Newton's law of inertia). 
The relativistic dynamics of the elementary particles, however, are not sufficient to describe completely nature: we must consider the undulatory nature of elementary particles. It is an experimental evidence that every elementary particle is characterized by a recurrence in time and space, determined by the kinematical state of the particle itself through the Planck constant $h$.  That is, to the energy and momentum of the particles are associated intrinsic temporal and spatial periodicities $T_t (\bar{\mathbf{p}})=h/\bar E (\bar{\mathbf{p}}) $ and $\lambda^i=h/p_i$.  Undulatory mechanics allows us to interpret  the energy-momentum of a particle and its space-time recurrence as two faces of the same coin. They are dual descriptions of the kinematical state. The instantaneous temporal and spatial periodicities of a particles can be written as a contravariant tangent four-vector $T^{\mu}=\{T_{t},\vec \lambda_{x}\}$, \cite{Dolce:2009ce}.  
 Indeed every particle can be formally represented by phasors or  waves, in which the space-time coordinates enter as  angular variable with periodicity $T^{\mu}$.  For instance, a free relativistic bosonic particle of four-momentum $\bar p_{\mu}$ can be described by a corresponding mode of a Klein-Gordon field, i.e. a standing wave of corresponding recurrence $T^{\mu}$. 

Since every system in nature is described in terms of elementary particles and every elementary particle is a ``periodic phenomenon'', it follows naturally that physics can be consistently described in terms of space-time ECs. To obtain a consistent relativistic description however, it is important to consider that  variations of kinematical state during interactions or changes in reference frame correspond to modulations of these ECs. For instance, we may think to the relativistic Doppler effect describing the modulations of temporal periodicity associated to Lorentz transformations or to the time dilatations in gravitational potentials. Actually, as note by de Broglie, the time and spatial recurrences of elementary particles (e.g. neutral bosons) are fully determined by their recurrences in the rest frame $T_{\tau}$, exactly as the energy and momentum are determined by Lorentz transformations from the mass of the particle. In fact the proper time recurrence is the Compton time $T_{\tau} = h / \bar M c$. The space-time periodicity in a given reference frame is therefore determined from this through Lorentz transformations: $c T_\tau= c \gamma(\bar{\mathbf{p}}) T_{t}(\bar{\mathbf{p}}) - \gamma(\bar{\mathbf{p}}) \vec \beta(\bar{\mathbf{p}}) \cdot \vec \lambda_{x}(\bar{\mathbf{p}})$. Considering the relativistic relations $\bar E(\bar{\mathbf{p}})  = \gamma(\bar{\mathbf{p}}) \bar M c^{2}$ and $\bar{\mathbf{p}} = c \vec \beta(\bar{\mathbf{p}})  \gamma(\bar{\mathbf{p}}) \bar M$, this confirms that $T^{\mu}$  is a four-vector fixed by the four-momentum ${\bar{p}}_{\mu} $: the condition is called de Broglie phase harmony $T_\tau \bar M c^2 \equiv T^{\mu} {\bar{p}}_{\mu}  \equiv h$. It is important to point out that relativity only describes the differential structure of space-time, without giving any explicit prescription about the boundary conditions. The assumption of intrinsic periodicity is in fact consistent with the variational principle applied to relativistic field theories. From this follows that a relativistic theory of ECs is a fully consistent relativistic theory. For the sake of simplicity, in this conceptual discussion we consider only time periodicity (through the equations of motion, this is sufficient to determine the spatial periodicity once the mass, i.e. the proper time periodicity, is known).  

A description of nature in terms of ECs has remarkable conceptual and physical implications \cite{Dolce:2009ce,Dolce:tune,Dolce:AdSCFT}. From a conceptual point of view, it implies that every particle can be regarded as a reference clock: the so-called ``internal clock''.  To see this we may consider the definition of  the unit of time in the international system (SI): a ``second'' is the duration of 9,192,631,770 characteristic cycles [...] of the Cs atom. The point that we want to stress is that time can be only defined  by counting the number of cycles of a phenomenon which is \emph{supposed} to be periodic --- or equivalent methods.  The assumption of intrinsic periodicity of a given phenomenon is crucial (and tautological) in the definition of time: since (as far as we know) we cannot travel in time, intrinsic periodicity guarantees that the unit of time does not change in different instants. Galileo, by comparing the librations of the lamp in the Pisa dome with his heart bits (``the human internal clock''), inferred the pendulum isochronism. But even more important he realized that such a persistent periodicity in time could be used as sufficiently accurate reference clock, allowing him to test the law of classical kinematics. 
The importance of the assumption of intrinsic periodicity in the definition of time is also evident in Einstein's definition of a relativistic clock: in a clock ``all that happens in a  given period is identical with all that happens in an arbitrary period''. Thus, an isolated elementary particle, owing its persistent periodicity, can be regarded as a reference clock, the so-called de Broglie ``internal clock'' \cite{Ferber:1996}. Every isolated particle (constant energy) has a persistent recurrence in time as an isolated pendulum. Therefore, similarly to the atomic clocks, by counting the number of its ``ticks'' it is possible to use it  to define time (note that similarly to elementary free particles internal clock, the characteristic periodicity of the atomic clock is fixed by the energy gaps in the electronic structure of the atom, through $h$). However, even a light particle such as the electron has an extremely fast intrinsic periodicity of $T_{\tau } = 8.093299724 \pm 11 \times 10^{-21} s$ (Compton time)  with respect to the periodicity, e.g., of the Cs clock which is of the order of $10^{-10} s$ . The observation of the internal clock of the elementary particles is more than a gedankenexperiment, as recent experiments are actually approaching this goal \cite{2008FoPh...38..659C}. Therefore a reference clock defined with the electron ticks would bring a revolutionary improvement of many orders of magnitude in our time resolution. Indeed, the difference of time scale between the ``ticks'' of a Cs clock and of an electron is comparable with the difference between the age of the universe and the duration of a solar year.  

\vspace{-1em}  
\section{Relativistic time flow}

An isolated system or universe composed by a single particle has a purely cyclic evolution. In fact, every elementary particle is a ``periodic phenomenon'' with persistent time recurrence.  There are not  energy variations in this elementary system. Every period cannot be distinguished by the other.  The whole physical information of such a cyclic universe is contained in a single period: ``all that happens in a  given period is identical with all that happens in an arbitrary period''. The evolution is parametrized by a cyclic time variable. We can now imagine to add other persistent  ``periodic phenomena'' into this universe, such as isolated particles and atomic clocks. From each of them the external time axis $t \in \mathbb R$ can be defined by counting the number of  theirs ``ticks''. Such an external time axis  can be regarded as an angular variable of infinite periodicity, such as that associated to the recurrence of a massless particles with  very low energy. Thus the cyclic variables describing the recursive evolutions of every elementary clock can be parametrized by a common time parameter $t$. The evolution of such a composite system  of non interacting ``periodic phenomena'' however is not cyclic. The combination of periodicities not rational each other  describes an ergodic system. Its orbits pass arbitrarily close to the initial phase-space point without never passing from it again (the definition of elementary systems could depend on the resolution).  
Now we imagine to turn on interactions among the elementary particles of this ergodic system.  The energy propagates according to the retarded relativistic potential (relativistic undulatory mechanics is in fact based upon relativistic waves). After a given time delay, this induces a retarded variation of periodic regime of the ECs involved in the interaction, dependent on the amount of energy exchanged. In this case it is possible to establish a ``before'' and an ``after'' with respect to these retarded events in time (in this case all that happens in a  given period is \emph{not} identical with all that happens in an arbitrary period). Hence we have relativistic causality and time ordering in terms of local and retarded modulations of periodicity of the internal clocks  \cite{Dolce:2009ce}.  This also means that physical systems composed by interacting elementary particles, are characterized  by chaotic evolutions rather than ergodic ones, according to our empirical observation of nature. 

In such a chaotic evolution, every value  (``instant'') of the external relativistic time axis $t \in \mathbb R$, defined by means of the atomic or ``particle'' clocks,  is characterized by a unique combination of the phases of all the ``internal clocks'' of the system. That is, events in time can be uniquely fixed by combining reference cycles ``ticks'', as in a stopwatch or in a calendar. In everyday life we in fact fix events in time in terms as combinations of reference cycles of years, months, days, hours, minutes, seconds, etc. These  are conventionally assumed to be rational each other (in particular in a sexagesimal base), but  they need regular  adjustments as they mimic natural cycles (e.g. Moon and Earth rotations) which form ergodic systems (or chaotic systems if we also consider interactions).  The combinations of relativistic ECs depend on the reference frame of the observer, according to relativistic simultaneity: every observer experiences a different ``present''. Hence, owing this unique combinatory description of phases associated to every instant in time, the external time axis can be in principle dropped. The time flow of the system can be in fact decomposed as modulations of ECs. 

Besides the many fundamental physical motivations, such a description of time flow  has important philosophical and anthropological justifications (the list would be too long to be mentioned here).  
This picture can be better understood if we consider the role of the mediator of interactions among particles (for simplicity's sake we consider only photons).  As mentioned before, massive particles  (expect neutrinos) have typically extremely fast periodicities. On the other hand, the time periodicity of  photons can vary from zero to infinity. In fact they have zero mass and therefore infinite proper time recurrence:  we say that light has a ``frozen rest clock'' $T_{\tau } \equiv \infty $. 
In a relational description of time, the long temporal and spatial periodicities of light provide the long scale temporal structure. Photons cycles provide  the reference temporal axis of the ordinary interpretation of relativity (emphatically non cyclic time coordinate), allowing a common parameterization of the elementary cyclic phenomena. That is,  such long space-time scales with respect to the typical periodicities of matter fields can be used as a reference upon which the ordinary relativistic structure of space-time can be built. In other words, the ``frozen'' clock of light (or of gravity) set the causal structure of the ECs of nature. 
The assumption of intrinsic periodicity enforces the local nature of relativistic time. The helicity (clockwise or anticlockwise) of the internal clocks is  arbitrary. Since the flow of time is given by the combination of the elementary clocks ``ticks'', the same physical evolution is equivalently described by inverting the helicity of all the elementary clocks of the system. Furthermore, as the external time axis can be in principle defined with reference to every single particle, the inversion of the internal clock of a particle does not imply the inversion of the arrow of time of the whole system (i.e. of the external time axis). The inversion of the helicity of a single internal clock however describes a different physical system, it in fact means to transform a particle to the corresponding antiparticle (as in Feynman's interpretation). 

%
%

\vspace{-1em}  
\section{The ``missing link'' of quantum mechanics}
\label{sec-1}

The technological possibility to replace atomic clocks with the internal clock of, say, an electron would open a new frontier in physics (similarly to what happened with other improvements in time resolutions, see Galileo example above). This could allow us to control the ``gears'' of QM.  The enormous rapidity of these recurrences in fact justifies a novel semi-classical interpretation of QM. If we assume that an elementary particle is constrained to have intrinsic periodicity, that particle is like a ``particle in box''. That is, the assumption of intrinsic periodicity is a quantization condition. This extends the wave-particle duality: every particle can be in fact represented as a one dimensional string vibrating with characteristic space-time periodicities determined by its kinematical state. The harmonics of such  vibrating strings represent the quantum excitations of the particles. The resulting theory is the relativistic generalization  of the theory of sound (allowing vibrations along the time dimension) in which elementary particles are the analogous of sound sources. Roughly speaking we give a timber to de Broglie waves \cite{Dolce:tune}.  It is possible to show that a statistical description of the vibrations associated to the ECs leads to a formal correspondence with ordinary QM, for both the canonical and Feynman formulations. For instance, as in a cylindric geometry, a periodic phenomenon can arrive to a given final configuration by passing through infinite classical paths characterized by different windings numbers. In this way it is possible to prove that the classical evolution of modulated ECs are actually described by the ordinary Feynman path integral. The correspondence can be interpreted in terms of 't Hooft determinism \cite{'tHooft:2001ar}: ``there is a close relationship between a particle moving very fast in a circle and a harmonics quantum oscillator'' --- the former can be statistically described as a fluid with unitary total density, see Born rule, the latter is the basic constituent of  quantum fields, see Klein-Gordon field. The periodicity of typical pure quantum systems (complete coherence) are extremely fast with respect to our resolution in time; for electrodynamical systems this is of the order of the \emph{zitterbewegung} ($10^{-21} s$) --- though coherent electromagnetic systems  (lasers, superconductors,  condesates, etc) can have a slower recurrence (the thermal noise destroys the quantum recurrence). A quantum system can be therefore imagined as a die rolling too fast with respect to our resolution in time, so that the outcomes can only be described  in a statistical way (with implicit Heisenberg's relation); an imaginary observed with infinite resolution in time would have no fun playing dice since  the outcomes could be always predicted (``God doesn't play dice'' A. Einstein). Intrinsic periodicity could represent the physical principle (``missing link'' \cite{Ferber:1996}) behind QM. The statistical description of ECs matches quantum dynamics without involving any local hidden variable. This suggests a possible deterministic description of physics. 


%
%
%
\vspace{-1em}  
\section{Gauge interaction as relativistic clock modulation}
\label{sec-1}
 A formulation of physics in terms of space-time ECs implements undulatory mechanics (e.g. the wave particle duality) in the geometrodynamcis of the space-time dimensions. Every interaction can be equivalently described as modulations of the periodicity of the ECs, i.e. as modulations of the internal clocks of the particles. As known from general relativity, the modulations of (space-time) clocks can be equivalently encoded in corresponding deformations of the underlying space-time metric. In this way we have proved \cite{Dolce:tune} that gauge interactions (e.g. electromagnetism), as gravitational interaction, can be associated to space-time geometrodynamics, similar to original Weyl's proposal. The idea is to describe  the oscillatory motion of a particle interacting, say, electromagnetically in terms of corresponding local variations of reference frame, i.e. similarly to the description of a particle interacting gravitationally in general relativity. In this case (linear approximation) the transformations associated to gauge interactions are local transformations of flat reference frames. Thus, such a formulation of physics in terms of ECs points out an important relationship between gauge and gravitational interactions \cite{Dolce:tune}. 
 
\vspace{-1em}  
\section{Conclusion}
\label{sec-1}
The flow of time in physical systems can be decomposed in ECs. This possibility is provided by QM, which associates to every elementary constituent of matter a recurrence in time. Free elementary particles can be regarded as reference clocks for a more accurate operational definition of the time unit. A formalization of elementary particles as ECs provides natural solutions to controversial questions of physics as well as a novel concept of relativistic time flow \cite{Dolce:spatimecyles,Dolce:2009ce,Dolce:tune,Dolce:AdSCFT}.

\end{document}